\documentclass[prl,reprint,superscriptaddress,amsmath,amssymb,aps,showpacs, floatfix,]{revtex4-1}

\usepackage{graphicx}
\usepackage{dcolumn}
\usepackage{bm}
\usepackage{epstopdf}
\usepackage{color}
\usepackage{booktabs}
\usepackage{comment}
\usepackage{psfrag}

\def\openone{\leavevmode\hbox{\small1\kern-3.3pt\normalsize1}}

\begin{document}

\title{Room temperature Rydberg Single Photon Source}

\author{M. M. M\"uller}
\affiliation{Institut f\"ur Quanteninformationsverarbeitung, 
Universit\"at Ulm, 89081 Ulm, Germany}

\author{A. K\"olle}

\author{R. L\"ow}

\author{T. Pfau}
\affiliation{5. Physikalisches Institut, Universit\"at Stuttgart, 
70550 Stuttgart, Germany}

\author{T. Calarco}\affiliation{Institut f\"ur Quanteninformationsverarbeitung, 
Universit\"at Ulm, 89081 Ulm, Germany}

\author{S. Montangero}
\affiliation{Institut f\"ur Quanteninformationsverarbeitung, 
Universit\"at Ulm, 89081 Ulm, Germany}

\date{\today}
\pacs{xxx}

\begin{abstract}
We present an optimal protocol to implement a room temperature Rydberg single photon source within an experimental setup based on micro cells filled with thermal vapor. The optimization of a pulsed four wave mixing scheme allows to double the effective Rydberg blockade radius as compared to a simple Gaussian pulse scheme, releasing some of the constrains on the geometry of the micro cells. The performance of the optimized protocol is improved by 
about 70\% with respect to the standard protocol. 
\end{abstract}

\maketitle

The quest to find the perfect hardware to support quantum information processing is one of the main open problems to be solved to bridge the distance between  quantum information theory and technological applications. Indeed the conditions to be met are very demanding and can be summarized in the need of having a (many-body) quantum system perfectly isolated from the environment but under perfect control at will. One of the promising candidates to fulfill such stringent conditions are Rydberg atoms as their interaction strength can be tuned by twelve orders of magnitude by simply exciting them~\cite{SaffmanReview}. This very peculiar property is very difficult to be found in other architectures, as interaction is usually either very weak or always-on, and allowed for a fast development of their application in quantum information: phase gates~\cite{GaetanNatPhys09,IsenhowerPRL10} and recently a single photon source based on Rydberg excitation blockade~\cite{Kuzmich} have been demonstrated using ultracold 
atoms. Despite 
these exciting successes, the need for ultracold temperatures still limits their applicability and thus recent efforts have been made towards replacing ultracold atoms with thermal vapor \cite{Loew}. Indeed  there has been considerable progress toward similar results, demonstrating appropriate coherence times~\cite{Kuebler}, four wave mixing~\cite{Koelle} and van-der-Waals interaction~\cite{inprep}.  Thermal vapor cells have been shown to act as a long lived quantum memory \cite{Budker} as well as a probabilistic single photon source \cite{Kuzmich}. Thermal vapor cells have even been shown to support entanglement between quantum states of the spin variables in two distant cells \cite{Polzik}.
A global approach that combines quantum memories, single photon sources, quantum logic operations and efficient detectors on one platform is highly desirable e.g. to implement quantum repeater protocols like the one proposel by  DLCZ \cite{DLCZ}.

Here we propose a scheme to obtain a single photon source using a thermal vapor of Rubidium 87 and we study theoretically its feasibility. 
The proposal stems from previous works originally thought for ultracold  atoms where the Rydberg blockade has been exploited to prepare 
a collective single excitation state (a W-state) as a resource to accomplish different tasks~\cite{SaffmanWalker,lukin}. When the W-state decays it will emit a photon with directionality given by the imprinted laser phases. This decay process has been widely studied by \cite{Scully, Pedersen, Manassah,Dudin,Bariani} leading to the experimental realization of an ultracold single photon source in \cite{Kuzmich}. We exploit optimal control theory to achieve the best possible preparation of the W-state under the experimental constraints in the more challenging regime at room temperature. Quantum optimal control theory has been applied successfully to solve this class of problems in few-body quantum systems in many different setups~\cite{Brif} and recently it has been proven to be successful also to optimally drive many-body quantum systems dynamics~\cite{Doria, Caneva}.
We exploit this recently introduced technique to optimize the state preparation, simulating the unitary dynamics of the atomic ensemble driven 
by two lasers by means of the time-dependent Density Matrix Renormalization Group in its Matrix Product State (MPS) formulation that allows to include long-range interactions~\cite{schollwoeck, dummies, MPS}. This is the first application of optimal control theory to many-body quantum systems with long range interactions.
\begin{figure}
\includegraphics[width=0.45\textwidth]{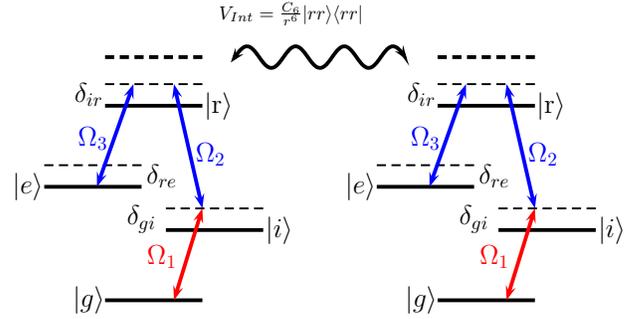}
\caption{Level Scheme of $^{87}$Rb and sketched interaction. The used states are $|g\rangle=5^2S_{1/2}$,  $|i\rangle=5^2P_{3/2}$, $|r\rangle=44D$, $|e\rangle=5^2P_{1/2}$. Interaction occurs between Rydberg ($|r\rangle$) states with potential $C_6/r^6$ and $C_6=h\cdot 5880\,\mathrm{MHz\, \mu m^6}$ \cite{Singer}.}\label{fig:level-scheme}
\end{figure}
We start our analyses with an ideal scenario of a frozen ordered unidimensional system: under such conditions, optimal control theory allows to find the best pulse to reach the W-state with very high fidelity. We then extract the general features of the found optimal pulses using them as initial guesses to drive random instances of atoms distributed within a three-dimensional sphere and optimize them. The resulting pulses provide a robust strategy to prepare the desired state despite the new disordered geometry and system dimensionality. Finally,  we study in details the effects of finite temperature (e.g. Doppler shifts) on the properties of the single-photon source and we show that we obtain a room temperature single 
photon source with performances similar to those obtained in the cold atoms case.

\paragraph*{System setup -}
We consider a hot rubidium vapor in a wedge shaped micro cell as recently introduced in \cite{Kuebler,Koelle}. At room temperature we can expect a linewidth of about $2\,$MHz for the 44D-state \cite{Adams}. Fig. \ref{fig:level-scheme} shows the relevant levels of the $^{87}$Rb atoms, where
$\Omega_i$ are the Rabi frequencies of the lasers and $\delta_{gi}$, $\delta_{ir}$ and $\delta_{re}$ are the detunings of the corresponding transitions.
By detuning $\Omega_1$ from the $\vert g\rangle \leftrightarrow \vert i\rangle$ transition we can adiabatically eliminate the level $\vert i\rangle$ in our theoretical model assuming $\delta_{ig}\gg \Omega_1,\Omega_2,\delta_{ir}$~\cite{Lambdasystem}.
From now on we consider the three level system $\vert g\rangle$, $\vert r\rangle$ and $\vert e\rangle$ and two driving lasers with Rabi frequencies $\Omega_{gr}= -\frac{\Omega_1\Omega_2}{2\delta_{gi}}$ and $\Omega_{re}= \Omega_3$.
The density of the vapor can be varied by changing the gas temperature. At $220^\circ\,$C this gives a density of $^{87}$Rb ($27.83\,\%$ of all Rubidium atoms) of $543\,\mathrm{atoms/\mu m^3}$.
The velocity distribution is Gaussian, corresponding to a thermal distribution. The decay rates are  $\tau_1=26.2\,\mathrm{ns}$ for the $\vert i\rangle \leftrightarrow \vert g\rangle$ transition, and  $\tau_2=27.7\,\mathrm{ns}$ for the $\vert e\rangle \leftrightarrow \vert g\rangle$ transition
\cite{Steck}. As the single photon shall be emitted by the latter transition, this sets also the time scale for the decay process.
This also limits the duration of the exciting laser pulses, that we set to be $T_0=2.5\,$ns. Given the separated timescales with respect to the observed coherence times of about a hundred ns, we simulate a closed, coherent system as far as the state preparation is concerned and analyze the non-coherent part of the system evolution separately. 
Finally, the Hamiltonian of the system $H=\sum_i H_{loc,i}+H_{int}$ after the adiabatic elimination and choosing $\Omega_1=\Omega_2$ is given by
\begin{align}
H_{loc,i}&=\frac{\Omega_{gr}}{2} (|g\rangle\langle r| +|r\rangle\langle g|) +\frac{\Omega_3}{2} (|r\rangle\langle e| +|e\rangle\langle r|)\nonumber\\
&+\delta_{ir} |r\rangle\langle r|+\delta_{re}|e\rangle\langle e|\,\\
 H_{int}&=\sum_{i\neq j}\frac{C_{6}}{r_{ij}^6}\vert r_{i}r_{j}\rangle\langle r_{i}r_{j}\vert.
\end{align}
If we change the pulse duration from $T_0$ to $T$ and at the same time we scale the Rabi frequencies and detunings by $T_0/T$, while typical length scales like the diameter $L_0$ of the system scale by $L_T=L_0(T/T_0)^{1/6}$ the Hamilton practically remains the same.

\paragraph*{State preparation -}
The goal of the manipulation protocol is to prepare a collective single excitation state 
$|W\rangle=\frac{1}{\sqrt{N}}\sum_{i=1}^N \mathrm{e}^{\imath\vec{k}_0\vec{r}_i}|e_i \rangle$
(where $|e_i\rangle=|g...geg...g\rangle$ means atom $i$ in state $e$ and all other atoms in state $g$), starting from the system's ground state $|\psi(0)\rangle= | gg \dots gg\rangle$. %
\begin{figure}
  \includegraphics[width=0.45\textwidth]{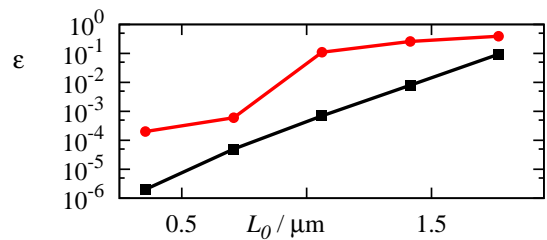}
 \caption{
  Final infidelity $\epsilon$ for a Gaussian $\pi$-pulses (circles) and optimal pulses (squares) in a one-dimensional $N$ atoms chain with lattice constant $a_0 = 0.35\,\mathrm{\mu m}$ at pulse duration $T_0=2.5\,\mathrm{ns}$ and total extension $L_0=a_0(N-1)$.}
 \label{chain}
\end{figure}%
In order to engineer the preparation of the W-state we simulate the time evolution of the system and optimize the pulse sequence. The time evolution is obtained by a tensor-network algorithm including two-body long-range interactions~\cite{schollwoeck,MPS}. The optimization is performed by means of the Chopped RAndom Bases (CRAB) optimal control technique for many-body quantum systems dynamics recently introduced in~\cite{Doria,Caneva}. 
The method is based on the a priori reduction of the problem complexity via a proper truncation of the accessible space used to describe the system wave function and the control field. In particular, in the system introduced above, the Rabi frequencies of two driving lasers $\Omega_{gr}(t)$ and $\Omega_{er}(t)$ are subject to optimization by the CRAB optimization method. The pulses are expanded in a truncated basis $\Omega_i=\sum_{j=1}^{N_i}c_{ij}f_{ij}(t)$ where the $c_{ij}$ are the coefficients of the expansion and the basis functions $f_{ij}(t)$ are chosen according to physical properties of the system. In particular here we choose the first principal harmonics of the Fourier expansion. The figure of merit to be minimized is the final infidelity of the system state evolved under the pulses $\Omega$ with respect to the target state $|W\rangle$, namely
\begin{equation}
\label{fom}
\epsilon= 1-\vert\langle W \vert \psi (T)\rangle\vert^2.
\end{equation}
The multi-variable function defined by (\ref{fom}), $\epsilon \equiv \epsilon(c_{ij})$ is then minimized by means of direct-search methods~\cite{Doria}. 
Typical parameters are $N_i=14$, $10^4$ iterations with $10^4$ Trotter steps and bound dimension $10:40$ for the MPS simulation.

\paragraph*{Results -}
We first concentrate on a one-dimensional chain of $N$ Rydberg atoms with fixed positions. Despite its simplicity 
-- as we shall show later -- this model already captures the essential features of the system dynamics as the long range interactions effectively map this system in a particular instance of the full 3D scenario that we consider later. An experimental realization can be with optical lattices like in \cite{arimondo}. The interaction between two excited Rydberg atoms can act as a blockade for double excitations and it enhances the driving Rabi frequency for the $\vert g\rangle \leftrightarrow \vert r\rangle$ transition by a factor of $\sqrt{N}$  in an ensemble of $N$ atoms, as one excitation suppresses $N-1$ other possible excitations  \cite{GaetanNatPhys09, BlockadeRadius, lukin}. 
\begin{figure}
  \includegraphics[width=0.48\textwidth]{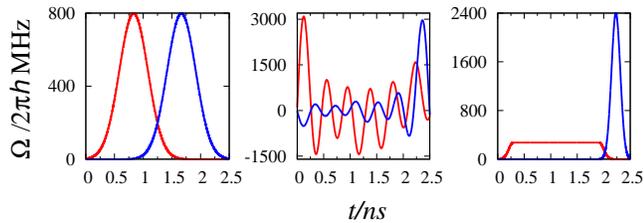}
   \caption{Left: Gaussian $\pi$-pulses guess. Middle: Pulse from optimization of 4 atoms chain. Right: Guess optimized for 3D cloud. Red: $\Omega_{gr}$, blue: $\Omega_{re}$.}\label{fig:pulses}
\end{figure}
Fig. \ref{chain} shows the final infidelity $\epsilon$ with respect to the $|W \rangle$ state obtained via Gaussian $\pi$-pulses (Fig. \ref{fig:pulses}, left) of atoms in a 1D chain with lattice spacing $a_0=0.35\,\mathrm{\mu m}$ for $T_0 =2.5\,\mathrm{ns}$ and total extension $L_0=a_0(N-1)$. Notice that for different numbers of atoms $N$ the Rabi frequency $\Omega_{gr}$ has  been rescaled by $1/\sqrt{N}$ to correct for the $\sqrt{N}$ enhancement due to the Rydberg blockade effect. From the figure it is clear that the Gaussian $\pi$-pulses fail for $L_0>0.7\,\mathrm{\mu m}$ (corresponding to more than three atoms) where $\Omega_{gr}$ gets comparable to the interaction strength. 
On the contrary the optimized pulses yield infidelity $\epsilon<10^{-2}$ up to $L_0\approx 1.4\,\mathrm{\mu m}$, thus allowing to almost double the system size.
So by optimization we can outperform the blockade radius associated with the guess pulse. This effect exceeds what one can expect by just stretching $\Omega_{gr}$ over the whole operation time and thus lowering its bandwidth. The presented pulses have only real Rabi frequencies. Allowing also for complex values (via phase modulation) roughly results in a 15\% relative improvement of the state preparation error.

\begin{figure}
   \includegraphics[width=0.45\textwidth]{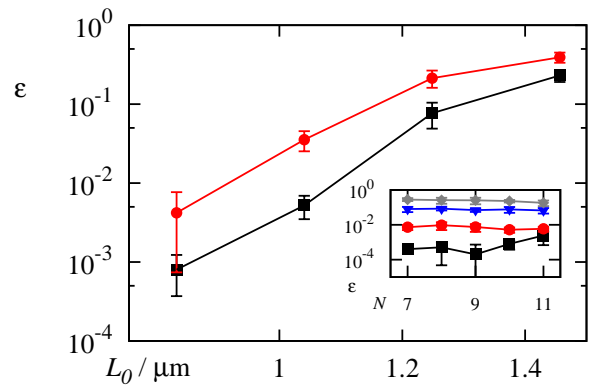}
  \caption{3D cloud of 10 randomly distributed atoms: infidelity $\epsilon$ as a function of the cloud's diameter $L_0$ plotted for Gaussian $\pi$-pulses (circles) and an optimized sequence (squares). The inset shows that $N$ has almost no influence on $\epsilon$. The curves in the inset correspond to $L_0=0.83\dots 1.46\,\mathrm{\mu m}$ and the optimized pulse.}\label{fig:3Dcloud}
\end{figure}

As stated in the introduction, our final goal is to prepare an ensemble of atoms with (uniformly distributed) random positions in 3D space moving randomly with thermally distributed velocities in the W-state. We thus now consider a 3D cloud of $N$ frozen atoms at random positions. Differently from the previous case, the interactions between atoms are random due to the distance-dependent interaction terms and this means that we have to produce very robust pulses ``on average" at the cost of some fidelity loss for each given sample of atoms. To this aim, we extract the relevant features of the optimized pulses e.g. Fig.~\ref{fig:pulses} (middle) and optimize only few parameters like the height and the width of the two pulses. The resulting optimal pulses are reported in the right panel of Fig.~\ref{fig:pulses}; improvement here is mainly due to the reduction of the bandwidth. The $\vert g\rangle \leftrightarrow \vert r\rangle$ transition is performed via a long flat pulse followed by a fast 
kick in the $\vert r\rangle \leftrightarrow \vert e\rangle$ 
transition.
Finally, we compare the performance of the two schemes: the infidelity $\epsilon$ obtained by W-state preparation with the optimized pulse and with the guess Gaussian $\pi$-pulses averaged over different system samples (i.e. different instances of the random positions). Fig.~\ref{fig:3Dcloud} shows the infidelity $\epsilon$ for clouds of ten atoms as a function of the maximum cloud extension $L_0$ 
while in the inset of Fig.~\ref{fig:3Dcloud}
$\epsilon$ is shown as a function of the number of atoms $N=7,8, \dots 11$ in the sample. In both cases the errorbars correspond to the statistical noise from 8 realizations of random atoms positions within a cloud of fixed maximum diameter $L_0$. The optimized pulses clearly result in improved infidelities and most importantly with almost no dependency on the number of atoms.  This result, as we shall show below, is corroborated also by our analytical estimate of accuracy of the optimal state preparation.
An additional complexity aspect for the experimental realization of the proposed protocol arises from the fact that the number $N$ of atoms in the ensemble depends on the density and the excitation volume and can only be determined with Poisson precision ($\pm \sqrt N$) changing a $\pi$-pulse roughly as $\cos(\frac{\pi}{2}\sqrt{1\pm N})\approx 1-\frac{\pi^2}{32N}$. This very weak dependence on the atom number $N$ of the final fidelity paves the way to a successful experimental realization of the W-state at room temperature.

We can provide an analytical estimate of the state preparation error under the hypothesis that: a) the major deviation of the prepared state from the desired W-state is given by the population in the two-excitations sector, b) each Rydberg atom is approximately performing an independent Rabi oscillation between $|g\rangle$ and $| r\rangle$ and c) the influence on an atom by another nearby excited one is to detune the single atom dynamics.  The detuning will just be the van-der-Waals interaction between two Rydberg excited atoms $V=C_6 d_{ij}^{-6}$, with $d_{ij}$ the distance between atom $i$ and atom $j$ and thus the resulting Rabi oscillation for this single atom is then only going up to having a fraction of $\frac{\Omega_{gr}^2}{\Omega_{gr}^2+V^2}=1-\frac{1}{1+c^{2}d_{ij}^{12}}$ in the excited (Rydberg) state ($c=\Omega_{gr}/C_6$). This double excitation part is missing in the population $P_i$ representing the first excitation sector. Summing this up over all neighbors gives
\begin{figure}
  \includegraphics[width=0.45\textwidth]{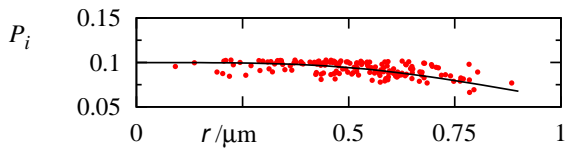}
 \caption{3D cloud of $N=10$ randomly distributed atoms: Errors in the state preparation. $P_i$ calculated by MPS simulation is compared to the theory curve from the $P(r)$ model.}\label{3Dcloud-asquared}
\end{figure}
\begin{equation}
 P_i \approx \frac{1}{N(N-1)}\sum_{j=1,j\neq i}^N \frac{1}{1+c^2d_{ij}^{12}}\,.
\end{equation}
In the thermodynamical limit an analogous continuous expression can be obtained by considering a homogeneous spherical cloud of radius $R$, 
introducing $P(r)$ as the single atom excitation. This yields
\begin{equation}\label{eq:asquared}
 P(r)\approx \frac{1}{NV}\int \frac{\rho^2\sin^2\theta \,d\rho\,d\theta\,d\phi\,}{1+c^2(r^2+\rho^2-2r\rho\sin\theta\cos\phi)^6}
\end{equation}
with $r$ the radial position (distance from the center of mass of the cloud) of atom $i$ in the cloud and integrating over the sphere's volume.
Fig. \ref{3Dcloud-asquared} shows the comparison between Eq.~(\ref{eq:asquared}) and the $P_i$ obtained from a numerical simulation of the dynamics of ten atoms. The good correspondence of this high density limit theory and numerical results of the few body simulation supports the previous finding of the very weak dependence of the results on the number of atoms $N$, that is, considering only ten atoms already is enough to get the major features of the state preparation. 

If we now include in our theoretical description the fact that the atoms are moving with a thermal distribution, we shall consider the lasers' Doppler shifts depending on the velocity of the atoms. Furthermore we shall analyze the effects of a time-dependent interaction $V(t)$ and the fact that the atoms might move out of the laser beam during state preparation, as at room temperature an average atom moves about $0.5\,\mathrm{\mu m}$ in $T_0$ time. For this time scale we can achieve good blockade within a sphere of radius $0.5-0.55\,\mathrm{\mu m}$ (corresponding to $L_0=1-1.1\,\mathrm{\mu m}$).
As the pulse is robust with respect to random positions we expect that the thermal motion  will not affect drastically the results.

Finally, the Doppler effect will cause a widening of the excitation populations with a Lorentzian shape $P_i^D\propto 1/(1+v_{i,\parallel}^2k^2/\Omega^2)$ depending on the velocity $v_{i,\parallel}$ of atom $i$ in the direction of the incoming lasers. The Doppler shift enters as an additional term $(k_1\pm k_2)v_{i,\parallel}|r\rangle\langle r|$ into $H_{loc,i}$ (we neglect the Doppler shift of the third laser since $\Omega_{re}\gg k_{re}v_i$). The wave numbers $k_i$ of lasers 1 ($\vert g\rangle \leftrightarrow \vert i\rangle$ transition) and 2 ($\vert i\rangle \leftrightarrow \vert r\rangle$ transition) are summed up or subtracted depending on whether the lasers are parallel or anti parallel.
We consider a sphere of radius $0.53\,\mathrm{\mu m}$ and an atomic density and velocity distribution corresponding to temperatures of $200-260\,^{\circ}$C, that is approximately $N=200-1200$ atoms and a velocity distribution characterized by a Gaussian width of $213-226\,\mathrm{m/s}$. 
We calculate the approximate W-state $|\tilde{W}\rangle=\sum_i\alpha_i|e_i\rangle$ (where the $\alpha_i$ contain the Doppler widening depending on the directions of the lasers) obtained by our state preparation pulses by simulating the evolution by $\Omega_{gr}$ for $N$ atoms assuming perfect blockade, thus keeping track only of the ground state and the $N$ singly excited states and constraining the system evolution to the ground state and the first excitation sector. To model the decay of $|\tilde{W}\rangle$ we follow the full exponential kernel description~ \cite{Pedersen,Eberly,Manassah,Scully,Dudin,Bariani}
and in addition we include also the motion of the particles.

 \begin{figure}
 \includegraphics[width=0.45\textwidth]{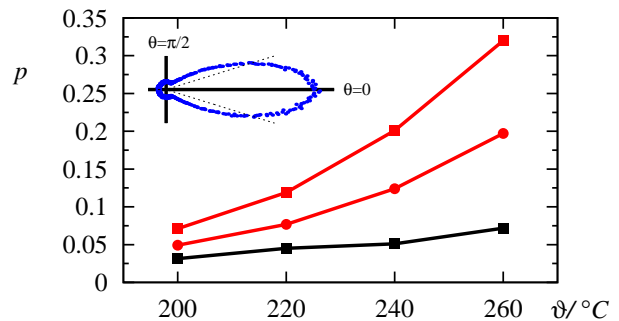}
 \caption{Directionality $p$ as a function of temperature $\vartheta$ for parallel lasers (red) and anti parallel lasers (black). Squares are values from state preparation with optimized pulses, circles from Gaussian $\pi$-pulses. The inset shows the emission cone on resonance for $\vartheta=220^{\circ}$C and anti parallel lasers.}\label{fig:photon}
\end{figure}

Fig. \ref{fig:photon}~ shows the resulting directionality $p$ (the probability of photon emission in the forward direction) as a function of the temperature, each value averaged over 240 random realizations of a uniform position distribution and Gaussian velocity distribution after a decay time of $100\,\mathrm{ns}$. The inset shows the emission cone on resonance for $\vartheta=220\,^{\circ}$C. The forward cone is defined to cover about $3\,\%$ of the solid angle corresponding to a maximum deviation from the forward direction of $0.3\,\mathrm{rad}$ (dashed lines). 
Values for $p$ are black for parallel and red for anti parallel lasers on the $\vert g\rangle \leftrightarrow \vert i\rangle$ and $\vert i\rangle \leftrightarrow \vert r\rangle$ transitions.
Clearly, the highest directionality is obtained for anti parallel lasers.
Squares correspond to state preparation with the optimized pulse. Circles mean state preparation with Gaussian $\pi$-pulses - in this case we have to decrease the cloud radius and thus deal only with about half as many atoms participating in the W-state leading to a reduction of the directionality.

At higher temperatures samples are denser and thus more atoms can be excited, resulting in higher directionality. However, the collision rate is enhanced thus coherence times are decreased.
Therefore the temperature has to be chosen properly. The best choice in the experimental settings is assumed to be in the range of $220^\circ\,$C, corresponding to an overall $70\,\%$ increase in directionality compared to the guess pulse.

\paragraph*{Conclusions -}
We have analyzed the possibility of constructing a single photon source using a thermal vapor of Rubidium 87. The results, however, are also qualitatively valid for other atoms like Cesium. By CRAB assisted pulse shaping using a Matrix Product State code we showed how to increase the number of atoms in a W-state by a factor of two compared to the initial guess which corresponds roughly to an increase of 70\,\% in directionality of the emitted photon. Choosing the exciting lasers as anti parallel gives an additional crucial increase. We plan to experimentally test these predictions also to quantify the effect of the collisions and laser focus not included in the present description and to see how they compare to experiments with ultracold atoms \cite{Kuzmich}.

\begin{acknowledgments}
The authors acknowledge support from  EU grants AQUTE and MALICIA; SFB/TRR21; QuOReP 
and we thank the bwGRiD project \footnote{%
    bwGRiD (www.bw-grid.de), member of the German D-Grid
    initiative,  funded by BMBF and MWK  Baden-W{\"{u}}rttemberg.}
  for the computational resources.
\end{acknowledgments}


\begin{thebibliography}{10}%
\makeatletter
\providecommand \@ifxundefined [1]{%
 \ifx #1\undefined \expandafter \@firstoftwo
 \else \expandafter \@secondoftwo
\fi
}%
\providecommand \@ifnum [1]{%
 \ifnum #1\expandafter \@firstoftwo
 \else \expandafter \@secondoftwo
\fi
}%
\providecommand \enquote [1]{``#1''}%
\providecommand \bibnamefont  [1]{#1}%
\providecommand \bibfnamefont [1]{#1}%
\providecommand \citenamefont [1]{#1}%
\providecommand\href[0]{\@sanitize\@href}%
\providecommand\@href[1]{\endgroup\@@startlink{#1}\endgroup\@@href}%
\providecommand\@@href[1]{#1\@@endlink}%
\providecommand \@sanitize [0]{\begingroup\catcode`\&12\catcode`\#12\relax}%
\@ifxundefined \pdfoutput {\@firstoftwo}{%
 \@ifnum{\z@=\pdfoutput}{\@firstoftwo}{\@secondoftwo}%
}{%
 \providecommand\@@startlink[1]{\leavevmode}%
 \providecommand\@@endlink[0]{}%
}{%
 \providecommand\@@startlink[1]{%
  \leavevmode
  \pdfstartlink
   attr{/Border[0 0 1 ]/H/I/C[0 1 1]}%
   user{/Subtype/Link/A<</Type/Action/S/URI/URI(#1)>>}%
  \relax
 }%
 \providecommand\@@endlink[0]{\pdfendlink}%
}%
\providecommand \url  [0]{\begingroup\@sanitize \@url }%
\providecommand \@url [1]{\endgroup\@href {#1}{\urlprefix}}%
\providecommand \urlprefix [0]{URL }%
\providecommand \Eprint[0]{\href }%
\@ifxundefined \urlstyle {%
  \providecommand \doi [1]{doi:\discretionary{}{}{}#1}%
}{%
  \providecommand \doi [0]{doi:\discretionary{}{}{}\begingroup
  \urlstyle{rm}\Url }%
}%
\providecommand \doibase [0]{http://dx.doi.org/}%
\providecommand \Doi[1]{\href{\doibase#1}}%
\providecommand \bibAnnote [3]{%
  \BibitemShut{#1}%
  \begin{quotation}\noindent
    \textsc{Key:}\ #2\\\textsc{Annotation:}\ #3%
  \end{quotation}%
}%
\providecommand \bibAnnoteFile [2]{%
  \IfFileExists{#2}{\bibAnnote {#1} {#2} {\input{#2}}}{}%
}%
\providecommand \typeout [0]{\immediate \write \m@ne }%
\providecommand \selectlanguage [0]{\@gobble}%
\providecommand \bibinfo [0]{\@secondoftwo}%
\providecommand \bibfield [0]{\@secondoftwo}%
\providecommand \translation [1]{[#1]}%
\providecommand \BibitemOpen[0]{}%
\providecommand \bibitemStop [0]{}%
\providecommand \bibitemNoStop [0]{.\EOS\space}%
\providecommand \EOS [0]{\spacefactor3000\relax}%
\providecommand \BibitemShut [1]{\csname bibitem#1\endcsname}%
\bibitem{SaffmanReview}%
  \bibfield{author}{%
  \bibinfo {author} {\bibfnamefont{M.}~\bibnamefont{Saffman}}, \bibinfo
  {author} {\bibfnamefont{T.~G.}\ \bibnamefont{Walker}},\ and\ \bibinfo
  {author} {\bibfnamefont{K.}~\bibnamefont{M\o{}lmer}},\ }%
  \bibfield{journal}{%
  \Doi{10.1103/RevModPhys.82.2313}{\bibinfo {journal} {Rev. Mod. Phys.}}\ }%
  \textbf{\bibinfo {volume} {82}},\ \bibinfo {pages} {2313} (\bibinfo {month}
  {Aug}\ \bibinfo {year} {2010}).
\bibitem{GaetanNatPhys09}%
  \bibfield{author}{%
  \bibinfo {author} {\bibfnamefont{A.}~\bibnamefont{Ga\"{e}tan}}, \bibinfo
  {author} {\bibfnamefont{Y.}~\bibnamefont{Miroshnychenko}}, \bibinfo {author}
  {\bibfnamefont{T.}~\bibnamefont{Wilk}}, \bibinfo {author}
  {\bibfnamefont{A.}~\bibnamefont{Chotia}}, \bibinfo {author}
  {\bibfnamefont{M.}~\bibnamefont{Vitaeu}}, \bibinfo {author}
  {\bibfnamefont{D.}~\bibnamefont{Comparat}}, \bibinfo {author}
  {\bibfnamefont{P.}~\bibnamefont{Pillet}}, \bibinfo {author}
  {\bibfnamefont{A.}~\bibnamefont{Browaeys}},\ and\ \bibinfo {author}
  {\bibfnamefont{P.}~\bibnamefont{Grangier}},\ }%
  \bibfield{journal}{%
  \Doi{10.1038/nphys1183}{\bibinfo {journal} {Nature Phys.}}\ }%
  \textbf{\bibinfo {volume} {5}},\ \bibinfo {pages} {115} (\bibinfo {year}
  {2009}).%
\bibitem{IsenhowerPRL10}%
  \bibfield{author}{%
  \bibinfo {author} {\bibfnamefont{L.}~\bibnamefont{Isenhower}}, \bibinfo
  {author} {\bibfnamefont{E.}~\bibnamefont{Urban}}, \bibinfo {author}
  {\bibfnamefont{X.~L.}\ \bibnamefont{Zhang}}, \bibinfo {author}
  {\bibfnamefont{A.~T.}\ \bibnamefont{Gill}}, \bibinfo {author}
  {\bibfnamefont{T.}~\bibnamefont{Henage}}, \bibinfo {author}
  {\bibfnamefont{T.~A.}\ \bibnamefont{Johnson}}, \bibinfo {author}
  {\bibfnamefont{T.~G.}\ \bibnamefont{Walker}},\ and\ \bibinfo {author}
  {\bibfnamefont{M.}~\bibnamefont{Saffman}},\ }%
  \bibfield{journal}{%
  \Doi{10.1103/PhysRevLett.104.010503}{\bibinfo {journal} {Phys. Rev. Lett.}}\
  }%
  \textbf{\bibinfo {volume} {104}},\ \bibinfo {pages} {010503} (\bibinfo
  {month} {Jan}\ \bibinfo {year} {2010}).%
\bibitem{Kuzmich}%
  \bibfield{author}{%
  \bibinfo {author} {\bibfnamefont{Y.~O.}\ \bibnamefont{Dudin}}\ and\ \bibinfo
  {author} {\bibfnamefont{A.}~\bibnamefont{Kuzmich}},\ }%
  \bibfield{journal}{%
  \Doi{10.1126/science.1217901}{\bibinfo {journal} {Science}}\ }%
  \textbf{\bibinfo {volume} {336}},\ \bibinfo {pages} {887} (\bibinfo {month}
  {May}\ \bibinfo {year} {2012}).%
\bibitem{Loew}\bibnamefont{R. L\"ow and T. Pfau}, Nature Photon. \textbf{3}, 197 News and Views (2009).
\bibitem{Kuebler}%
  \bibfield{author}{%
  \bibinfo {author} {\bibfnamefont{H.}~\bibnamefont{K\"ubler}}, \bibinfo
  {author} {\bibfnamefont{J.~P.}\ \bibnamefont{Shaffer}}, \bibinfo {author}
  {\bibfnamefont{T.}~\bibnamefont{Baluktsian}}, \bibinfo {author}
  {\bibfnamefont{R.}~\bibnamefont{L\"ow}},\ and\ \bibinfo {author}
  {\bibfnamefont{T.}~\bibnamefont{Pfau}},\ }%
  \bibfield{journal}{%
  \Doi{10.1038/nphoton.2009.260}{\bibinfo {journal} {Nat. Photon.}}\ }%
  \textbf{\bibinfo {volume} {4}},\ \bibinfo {pages} {112} (\bibinfo {month}
  {Feb}\ \bibinfo {year} {2010}).%
\bibitem{Koelle}%
  \bibfield{author}{%
  \bibinfo {author} {\bibfnamefont{A.}~\bibnamefont{K\"olle}}, \bibinfo
  {author} {\bibfnamefont{G.}~\bibnamefont{Epple}}, \bibinfo {author}
  {\bibfnamefont{H.}~\bibnamefont{K\"ubler}}, \bibinfo {author}
  {\bibfnamefont{R.}~\bibnamefont{L\"ow}},\ and\ \bibinfo {author}
  {\bibfnamefont{T.}~\bibnamefont{Pfau}},\ }%
  \bibfield{journal}{%
  \Doi{10.1103/PhysRevA.85.063821}{\bibinfo {journal} {Phys. Rev. A}}\ }%
  \textbf{\bibinfo {volume} {85}},\ \bibinfo {pages} {063821} (\bibinfo {month}
  {Jun}\ \bibinfo {year} {2012}).
\bibitem{inprep}%
  \bibfield{author}{%
  \bibinfo {author} {\bibfnamefont{T.}~\bibnamefont{Baluktsian}}, \bibinfo {author}
  {\bibfnamefont{B.}~\bibnamefont{Huber}}, \bibinfo {author}
  {\bibfnamefont{R.}~\bibnamefont{L\"ow}},\ and\ \bibinfo {author}
  {\bibfnamefont{T.}~\bibnamefont{Pfau}},\ }%
  \bibinfo {journal} {submitted.}%
\bibitem{Budker}\bibfnamefont{M. V. Balabas, T. Karaulanov, M. P. Ledbetter, and D. Budker}, Phys. Rev. Lett. \textbf{105}, 070801 (Aug 2010).
\bibitem{Polzik}\bibfnamefont{B. Julsgaard, A. Kozhekin, and Eugene S. Polzik}, Nature \textbf{413}, 400 (2001).
\bibitem{DLCZ}\bibfnamefont{L.-M. Duan, M. D. Lukin, J. I. Cirac, and P. Zoller}, Nature \textbf{414}, 413 (2001).
\bibitem{SaffmanWalker}%
\bibfield{journal}{%
    }%
  \bibfield{author}{%
  \bibinfo {author} {\bibfnamefont{M.}~\bibnamefont{Saffman}}\ and\ \bibinfo
  {author} {\bibfnamefont{T.~G.}\ \bibnamefont{Walker}},\ }%
  \bibfield{journal}{%
  \Doi{10.1103/PhysRevA.66.065403}{\bibinfo {journal} {Phys. Rev. A}}\ }%
  \textbf{\bibinfo {volume} {66}},\ \bibinfo {pages} {065403} (\bibinfo {month}
  {Dec}\ \bibinfo {year} {2002}).
\bibitem{lukin}%
  \bibfield{author}{%
  \bibinfo {author} {\bibfnamefont{M.~D.}\ \bibnamefont{Lukin}}, \bibinfo
  {author} {\bibfnamefont{M.}~\bibnamefont{Fleischhauer}}, \bibinfo {author}
  {\bibfnamefont{R.}~\bibnamefont{Cote}}, \bibinfo {author}
  {\bibfnamefont{L.~M.}\ \bibnamefont{Duan}}, \bibinfo {author}
  {\bibfnamefont{D.}~\bibnamefont{Jaksch}}, \bibinfo {author}
  {\bibfnamefont{J.~I.}\ \bibnamefont{Cirac}},\ and\ \bibinfo {author}
  {\bibfnamefont{P.}~\bibnamefont{Zoller}},\ }%
  \bibfield{journal}{%
  \Doi{10.1103/PhysRevLett.87.037901}{\bibinfo {journal} {Phys. Rev. Lett.}}\
  }%
  \textbf{\bibinfo {volume} {87}},\ \bibinfo {pages} {037901} (\bibinfo {month}
  {Jun}\ \bibinfo {year} {2001}).
\bibitem{Scully}%
  \bibfield{author}{%
  \bibinfo {author} {\bibfnamefont{M.~O.}\ \bibnamefont{Scully}}, \bibinfo
  {author} {\bibfnamefont{E.~S.}\ \bibnamefont{Fry}}, \bibinfo {author}
  {\bibfnamefont{C.~H.~R.}\ \bibnamefont{Ooi}},\ and\ \bibinfo {author}
  {\bibfnamefont{K.}~\bibnamefont{W\'odkiewicz}},\ }%
  \bibfield{journal}{%
  \Doi{10.1103/PhysRevLett.96.010501}{\bibinfo {journal} {Phys. Rev. Lett.}}\
  }%
  \textbf{\bibinfo {volume} {96}},\ \bibinfo {pages} {010501} (\bibinfo {month}
  {Jan}\ \bibinfo {year} {2006}).
\bibitem{Pedersen}%
  \bibfield{author}{%
  \bibinfo {author} {\bibfnamefont{L.~H.}\ \bibnamefont{Pedersen}}\ and\
  \bibinfo {author} {\bibfnamefont{K.}~\bibnamefont{M\o{}lmer}},\ }%
  \bibfield{journal}{%
  \Doi{10.1103/PhysRevA.79.012320}{\bibinfo {journal} {Phys. Rev. A}}\ }%
  \textbf{\bibinfo {volume} {79}},\ \bibinfo {pages} {012320} (\bibinfo {month}
  {Jan}\ \bibinfo {year} {2009}).
\bibitem{Manassah}%
  \bibfield{author}{%
  \bibinfo {author} {\bibfnamefont{R.}~\bibnamefont{Friedberg}}\ and\ \bibinfo
  {author} {\bibfnamefont{J.~T.}\ \bibnamefont{Manassah}},\ }%
  \bibfield{journal}{%
  \Doi{10.1016/j.physleta.2007.11.064}{\bibinfo {journal} {Physics Letters A}}\
  }%
  \textbf{\bibinfo {volume} {372}},\ \bibinfo {pages} {2514 } (\bibinfo {year}
  {2008}),\ ISSN \bibinfo {issn} {0375-9601}.
  \bibfield{author2}{%
  \bibinfo {author} {\bibfnamefont{R.}~\bibnamefont{Friedberg}}\ and\ \bibinfo
  {author} {\bibfnamefont{J.~T.}\ \bibnamefont{Manassah}},\ }%
  \bibfield{journal2}{%
  \Doi{10.1016/j.physleta.2007.12.056}{\bibinfo {journal} {Physics Letters A}}\
  }%
  \textbf{\bibinfo {volume} {372}},\ \bibinfo {pages} {2787 } (\bibinfo {year}
  {2008}),\ ISSN \bibinfo {issn} {0375-9601}.
  \bibfield{author3}{%
  \bibinfo {author} {\bibfnamefont{R.}~\bibnamefont{Friedberg}}\ and\ \bibinfo
  {author} {\bibfnamefont{J.~T.}\ \bibnamefont{Manassah}},\ }%
  \bibfield{journa3l}{%
  \Doi{10.1016/j.physleta.2008.09.032}{\bibinfo {journal} {Physics Letters A}}\
  }%
  \textbf{\bibinfo {volume} {372}},\ \bibinfo {pages} {6833 } (\bibinfo {year}
  {2008}),\ ISSN \bibinfo {issn} {0375-9601}.
  \bibfield{author4}{%
  \bibinfo {author} {\bibfnamefont{R.}~\bibnamefont{Friedberg}}\ and\ \bibinfo
  {author} {\bibfnamefont{J.~T.}\ \bibnamefont{Manassah}},\ }%
  \bibfield{journal4}{%
  \Doi{10.1016/j.optcom.2008.05.003}{\bibinfo {journal} {Optics
  Communications}}\ }%
  \textbf{\bibinfo {volume} {281}},\ \bibinfo {pages} {4391 } (\bibinfo {year}
  {2008}),\ ISSN \bibinfo {issn} {0030-4018}.
\bibitem{Dudin}%
  \bibfield{author}{%
  \bibinfo {author} {\bibfnamefont{Y.~O.}\ \bibnamefont{Dudin}}, \bibinfo
  {author} {\bibfnamefont{F.}~\bibnamefont{Bariani}},\ and\ \bibinfo {author}
  {\bibfnamefont{A.}~\bibnamefont{Kuzmich}},\ }%
  \bibfield{journal}{%
  \Doi{10.1103/PhysRevLett.109.133602}{\bibinfo {journal} {Phys. Rev. Lett.}}\
  }%
  \textbf{\bibinfo {volume} {109}},\ \bibinfo {pages} {133602} (\bibinfo
  {month} {Sep}\ \bibinfo {year} {2012}).
\bibitem{Bariani}%
  \bibfield{author}{%
  \bibinfo {author} {\bibfnamefont{F.}~\bibnamefont{Bariani}}, \bibinfo
  {author} {\bibfnamefont{Y.~O.}\ \bibnamefont{Dudin}}, \bibinfo {author}
  {\bibfnamefont{T.~A.~B.}\ \bibnamefont{Kennedy}},\ and\ \bibinfo {author}
  {\bibfnamefont{A.}~\bibnamefont{Kuzmich}},\ }%
  \bibfield{journal}{%
  \Doi{10.1103/PhysRevLett.108.030501}{\bibinfo {journal} {Phys. Rev. Lett.}}\
  }%
  \textbf{\bibinfo {volume} {108}},\ \bibinfo {pages} {030501} (\bibinfo
  {month} {Jan}\ \bibinfo {year} {2012}).
\bibitem{Brif}%
  \bibfield{author}{%
  \bibinfo {author} {\bibfnamefont{C.}~\bibnamefont{Brif}}, \bibinfo {author}
  {\bibfnamefont{R.}~\bibnamefont{Chakrabarti}},\ and\ \bibinfo {author}
  {\bibfnamefont{H.}~\bibnamefont{Rabitz}},\ }%
  \bibfield{journal}{%
  \bibinfo {journal} {New Journal of Physics}\ }%
  \textbf{\bibinfo {volume} {12}},\ \bibinfo {pages} {075008} (\bibinfo {year}
  {2010}).
\bibitem{Doria}%
  \bibfield{author}{%
  \bibinfo {author} {\bibfnamefont{P.}~\bibnamefont{Doria}}, \bibinfo {author}
  {\bibfnamefont{T.}~\bibnamefont{Calarco}},\ and\ \bibinfo {author}
  {\bibfnamefont{S.}~\bibnamefont{Montangero}},\ }%
  \bibfield{journal}{%
  \Doi{10.1103/PhysRevLett.106.190501}{\bibinfo {journal} {Phys. Rev. Lett.}}\
  }%
  \textbf{\bibinfo {volume} {106}},\ \bibinfo {pages} {190501} (\bibinfo
  {month} {May}\ \bibinfo {year} {2011}).
\bibitem{Caneva}%
  \bibfield{author}{%
  \bibinfo {author} {\bibfnamefont{T.}~\bibnamefont{Caneva}}, \bibinfo {author}
  {\bibfnamefont{T.}~\bibnamefont{Calarco}},\ and\ \bibinfo {author}
  {\bibfnamefont{S.}~\bibnamefont{Montangero}},\ }%
  \bibfield{journal}{%
  \Doi{10.1103/PhysRevA.84.022326}{\bibinfo {journal} {Phys. Rev. A}}\ }%
  \textbf{\bibinfo {volume} {84}},\ \bibinfo {pages} {022326} (\bibinfo {month}
  {Aug}\ \bibinfo {year} {2011}).
\bibitem{schollwoeck}%
  \bibfield{author}{%
  \bibinfo {author} {\bibfnamefont{U.}~\bibnamefont{Schollw\"ock}},\ }%
  \bibfield{journal}{%
  \Doi{10.1103/RevModPhys.77.259}{\bibinfo {journal} {Rev. Mod. Phys.}}\ }%
  \textbf{\bibinfo {volume} {77}},\ \bibinfo {pages} {259} (\bibinfo {month}
  {Apr}\ \bibinfo {year} {2005}).%
\bibitem{dummies}%
  \bibfield{author}{%
  \bibinfo {author} {\bibfnamefont{G.}~\bibnamefont{De~Chiara}}, \bibinfo
  {author} {\bibfnamefont{M.}~\bibnamefont{Rizzi}}, \bibinfo {author}
  {\bibfnamefont{D.}~\bibnamefont{Rossini}},\ and\ \bibinfo {author}
  {\bibfnamefont{S.}~\bibnamefont{Montangero}},\ }%
  \bibfield{journal}{%
  \Doi{doi:10.1166/jctn.2008.011}{\bibinfo {journal} {Journal of Computational
  and Theoretical Nanoscience}}\ }%
  \textbf{\bibinfo {volume} {5}},\ \bibinfo {pages} {1277} (\bibinfo {year}
  {2008}).
\bibitem{MPS}%
  \bibfield{author}{%
  \bibinfo {author} {\bibfnamefont{U.}~\bibnamefont{Schollw\"ock}},\ }%
  \bibfield{journal}{%
  \Doi{10.1016/j.aop.2010.09.012}{\bibinfo {journal} {Annals of Physics}}\ }%
  \textbf{\bibinfo {volume} {326}},\ \bibinfo {pages} {96 } (\bibinfo {year}
  {2011}),\ ISSN \bibinfo {issn} {0003-4916},\ \bibinfo {note} {january 2011
  Special Issue}.
\bibitem{Adams}%
  \bibfield{author}{%
  \bibinfo {author} {\bibfnamefont{A. K.}~\bibnamefont{Mohapatra}}, \bibinfo {author}
  {\bibfnamefont{T. R.}~\bibnamefont{Jackson}},\ and\ \bibinfo {author}
  {\bibfnamefont{C. S.}~\bibnamefont{Adams}},\ }%
  \bibfield{journal}{%
  \Doi{10.1103/PhysRevLett.98.113003}{\bibinfo {journal} {Phys. Rev. Lett.}}\
  }%
  \textbf{\bibinfo {volume} {98}},\ \bibinfo {pages} {113003} (\bibinfo
  {month} {Mar}\ \bibinfo {year} {2007}).
\bibitem{Singer}%
  \bibfield{author}{%
  \bibinfo {author} {\bibfnamefont{K.}~\bibnamefont{Singer}},\ }%
  \emph{\bibinfo {title} {Interactions in an ultracold gas of Rydberg atoms}},\
  Ph.D. thesis,\ \bibinfo {school} {Universit\"at Freiburg} (\bibinfo {year}
  {2004}).
\bibitem{Lambdasystem}%
  \bibfield{author}{%
  \bibinfo {author} {\bibfnamefont{E.}~\bibnamefont{Brion}}, \bibinfo {author}
  {\bibfnamefont{L.~H.}\ \bibnamefont{Pedersen}},\ and\ \bibinfo {author}
  {\bibfnamefont{K.}~\bibnamefont{M\o{}lmer}},\ }%
  \bibfield{journal}{%
  \bibinfo {journal} {Journal of Physics A}\ }%
  \textbf{\bibinfo {volume} {40}},\ \bibinfo {pages} {1033} (\bibinfo {year}
  {2007}).
\bibitem{Steck}%
  \bibfield{author}{%
  \bibinfo {author} {\bibfnamefont{D.~A.}\ \bibnamefont{Steck}},\ }%
  \enquote{\bibinfo {title} {Rubidium 87 d line data (revision 2.1.4, 23
  december 2010)},}\ \url{steck.us/alkalidata}.
\bibitem{arimondo}%
  \bibfield{author}{%
  \bibinfo {author} {\bibfnamefont{M.}~\bibnamefont{Viteau}}, \bibinfo {author}
  {\bibfnamefont{M.~G.}\ \bibnamefont{Bason}}, \bibinfo {author}
  {\bibfnamefont{J.}~\bibnamefont{Radogostowicz}}, \bibinfo {author}
  {\bibfnamefont{N.}~\bibnamefont{Malossi}}, \bibinfo {author}
  {\bibfnamefont{D.}~\bibnamefont{Ciampini}}, \bibinfo {author}
  {\bibfnamefont{O.}~\bibnamefont{Morsch}},\ and\ \bibinfo {author}
  {\bibfnamefont{E.}~\bibnamefont{Arimondo}},\ }%
  \bibfield{journal}{%
  \Doi{10.1103/PhysRevLett.107.060402}{\bibinfo {journal} {Phys. Rev. Lett.}}\
  }%
  \textbf{\bibinfo {volume} {107}},\ \bibinfo {pages} {060402} (\bibinfo
  {month} {Aug}\ \bibinfo {year} {2011}).
\bibitem{BlockadeRadius}%
  \bibfield{author}{%
  \bibinfo {author} {\bibfnamefont{R.}~\bibnamefont{Heidemann}}, \bibinfo
  {author} {\bibfnamefont{U.}~\bibnamefont{Raitzsch}}, \bibinfo {author}
  {\bibfnamefont{V.}~\bibnamefont{Bendkowsky}}, \bibinfo {author}
  {\bibfnamefont{B.}~\bibnamefont{Butscher}}, \bibinfo {author}
  {\bibfnamefont{R.}~\bibnamefont{L\"ow}}, \bibinfo {author}
  {\bibfnamefont{L.}~\bibnamefont{Santos}},\ and\ \bibinfo {author}
  {\bibfnamefont{T.}~\bibnamefont{Pfau}},\ }%
  \bibfield{journal}{%
  \Doi{10.1103/PhysRevLett.99.163601}{\bibinfo {journal} {Phys. Rev. Lett.}}\
  }%
  \textbf{\bibinfo {volume} {99}},\ \bibinfo {pages} {163601} (\bibinfo {month}
  {Oct}\ \bibinfo {year} {2007}).
\bibitem{Eberly}%
  \bibfield{author}{%
  \bibinfo {author} {\bibfnamefont{J.~H.}\ \bibnamefont{Eberly}},\ }%
  \bibfield{journal}{%
  \bibinfo {journal} {Journal of Physics B: Atomic, Molecular and Optical
  Physics}\ }%
  \textbf{\bibinfo {volume} {39}},\ \bibinfo {pages} {S599} (\bibinfo {year}
  {2006}).
\bibitem{Note1}%
  \BibitemOpen
  \bibinfo {note} {BwGRiD (www.bw-grid.de), member of the German D-Grid
  initiative, funded by BMBF and MWK Baden-W{\"{u}}rttemberg.}%
  \bibAnnoteFile{Stop}{Note1}%
\end{thebibliography}
\end{document}